# Electronic Raman-scattering study of low-energy excitations in single and double CuO$_2$-layer Tl-Ba-(Ca)-Cu-O superconductors

Moonsoo Kang
*Department of Physics and Science and Technology Center for Superconductivity, University of Illinois at Urbana-Champaign, Urbana, Illinois 61801*

G. Blumberg
*Department of Physics and Science and Technology Center for Superconductivity, University of Illinois at Urbana-Champaign, Urbana, Illinois 61801
and Institute of Chemical Physics and Biophysics, Rävala 10, Tallinn EE0001, Estonia*

M. V. Klein
*Department of Physics and Science and Technology Center for Superconductivity, University of Illinois at Urbana-Champaign, Urbana, Illinois 61801*

N. N. Kolesnikov
*Institute for Solid State Physics, Russian Academy of Science, Chernogolovka, Moscow Region, 142 432, Russia*

The low-energy Raman continuum and the redistribution of the continuum to a peak (the ''2Δ peak'') in the superconducting state have been studied in Tl-Ba-(Ca)-Cu-O superconductors with a single CuO$_2$ layer (Tl-2201) and a double CuO$_2$ layer (Tl-2212). The $2\Delta/k_BT_c$ ratios in $A_{1g}$ and $B_{1g}$ symmetries are larger for Tl-2212 than for Tl-2201. The $B_{1g}/A_{1g}$ gap ratio is also larger in Tl-2212. The $A_{1g}$ intensities of the continuum and the 2Δ peak are significantly weaker than the $B_{1g}$ intensities in Tl-2201, but are comparable in Tl-2212. This shows that the Coulomb screening is much stronger in Tl-2201. The change from Tl-2201 to Tl-2212 of the normalized $A_{1g}$ 2Δ peak intensity is identical within experimental error to that of the normalized $A_{1g}$ continuum intensity. This suggests that the excitations forming the 2Δ peak and the continuum couple to light by the same mechanism. [S0163-1829(97)51842-X]

In electronic Raman scattering experiments, excitations from different areas on the Fermi surface can be probed by selecting different polarization geometries of incident and scattered photons. Thus, electronic Raman scattering can give vital information such as the magnitude and anisotropy of the superconducting gap. In conventional superconductors, peaks from the quasiparticle excitations across the superconducting gap are observed in superconducting states,[1] and are well described by the existing theory.[2] In the case of cuprate superconductors, however, a flat, featureless, and hardly temperature-dependent electronic excitation (continuum) exists over a broad range of energy, and a broad peak (the 2Δ peak) and suppression of intensity below this peak are observed in the superconducting state. It has been proposed that the continuum comes from the incoherent scattering of the quasiparticles in strongly correlated systems,[3–5] but there is yet incomplete understanding of the continuum in the normal state. Theoretical studies of the 2Δ peak in cuprate superconductors reported thus far[6,7] are mainly based on the extension of the conventional theory[2] to the anisotropic gap, and fail to give a proper description of the continuum. However, as a result of a resonance study of electronic Raman scattering in Tl-2201, it was reported recently that the 2Δ peak comes from the redistribution of the continuum.[8] This implies that a proper treatment of the continuum is essential for a correct description of the electronic Raman scattering in high-temperature superconductors.

Tl-based high-temperature superconductors are important not only because they have high transition temperatures but because they give us a chance to study the effect of stacking multiple CuO$_2$ layers in the unit cell. Other families of materials, the Bi-based or Hg-based superconductors, have several structures with different numbers of CuO$_2$ layers, but sample quality and availability are somewhat limited for a complete study with these materials. From the Tl-Ba-(Ca)-Cu-O system with the general formula Tl$_2$Ba$_2$Ca$_{n-1}$Cu$_n$O$_{2n+4}$ ($n=1–3$), we have studied samples with a single CuO$_2$ layer (Tl$_2$Ba$_2$CuO$_6$; Tl-2201) and a double CuO$_2$ layer (Tl$_2$Ba$_2$CaCu$_2$O$_8$; Tl-2212).

In this communication, we report the electronic Raman scattering study of Tl-Ba-(Ca)-Cu-O superconductors with single and double CuO$_2$ layers. The magnitude and the anisotropy of the superconducting gap were measured from those materials. The relative scattering intensities in $A_{1g}$ and $B_{1g}$ symmetries are compared to examine the effect of the Coulomb screening on single and double CuO$_2$ layer materials.

The experiments were done on a Tl-2201 single crystal with $T_c=85$ K and a Tl-2212 single crystal with $T_c=102$ K grown as described in Ref. 9. The structures of Tl-2201 and Tl-2212 are very similar except for the number of neighboring CuO$_2$ planes. In both materials, single or double CuO$_2$ layers are separated by Ba-O and Tl-O layers, and there is a Ca atom between neighboring CuO$_2$ planes in Tl-2212.

Both materials have tetragonal unit cells with point group $D_{4h}(I4/mmm)$. The crystals have natural mirrorlike $ab$-plane surfaces and typical dimensions of $\sim 1\times 1\times 0.05$ mm$^3$. The transition temperatures were determined by magnetization measurements. The Raman spectra reported here were obtained in pseudo-backscattering geometry using a conventional (macro) Raman-scattering setup and a custom micro-Raman system which has an aberration-free low-temperature capability.[10] We used a high-energy (blue) excitation (2.73 eV) and a low-energy (red) excitation (1.92 eV) from a Kr$^+$ laser. The laser excitation was focused onto a 5-$\mu$m-diam. spot in the conventional setup and onto a 2 $\mu$m spot in the micro-Raman setup. The laser power was reduced to a level which does not increase the temperature of the illuminated spot significantly. The temperatures referred to in this communication are the nominal temperatures inside the cryostat. The spectra were taken by a triple grating spectrometer with a liquid nitrogen cooled CCD detector and corrected for the spectral response of the spectrometer and the detector.

The Raman-scattering intensity is given by the imaginary part of the Raman response function $\chi''(\omega)$ via the fluctuation-dissipation theorem,

$$I(\omega) \propto [1+n(\omega)]\chi''(\omega), \qquad (1)$$

where $n(\omega)=1/(e^{\omega/T}-1)$ is the Bose factor. Within a one-band model, labeled by wave vector $\mathbf{k}$, the Raman response function for quasiparticle excitations is given by

$$\chi''(\omega) = \mathrm{Im}\left[\sum_{\mathbf{k}}|\gamma_{\mathbf{k}}|^2 P_{\mathbf{k}}(i\omega) - \frac{(\Sigma_{\mathbf{k}}\gamma_{\mathbf{k}}P_{\mathbf{k}}(i\omega))^2}{\Sigma_{\mathbf{k}}P_{\mathbf{k}}(i\omega)}\right]_{i\omega\to\omega+i\delta} \qquad (2)$$

where $\gamma_{\mathbf{k}}$ is a light-scattering vertex function (or Raman-scattering form factor), $\delta$ is positive infinitesimal, and $P_{\mathbf{k}}(i\omega)$ is a frequency summed polarization bubble written as

$$P_{\mathbf{k}}(i\omega) = \frac{1}{\beta}\sum_{\omega'}\mathcal{G}(\mathbf{k}+\mathbf{q},i\omega'+i\omega)\mathcal{G}(\mathbf{k},i\omega'), \qquad (3)$$

where $\mathcal{G}(\mathbf{k},i\omega)$ is the Matsubara Green's function, $\beta$ is the inverse temperature, and the frequency summation is done over the Matsubara frequencies. In a conventional BCS-type superconductor, $P_{\mathbf{k}}(i\omega)$ can be easily calculated and becomes the Tsuneto function given by[18]

$$\lambda_{\mathbf{k}}(i\omega) = \frac{\Delta(\mathbf{k})^2}{E(\mathbf{k})^2}\tanh\left(\frac{E(\mathbf{k})}{2k_BT}\right)$$
$$\times\left[\frac{1}{2E(\mathbf{k})+i\omega} + \frac{1}{2E(\mathbf{k})-i\omega}\right], \qquad (4)$$

where $E(\mathbf{k})=\sqrt{\epsilon_{\mathbf{k}}^2+\Delta(\mathbf{k})^2}$ is the quasiparticle energy in the superconducting state, $\epsilon_{\mathbf{k}}$ is the normal quasiparticle energy minus the Fermi energy, and $\Delta_{\mathbf{k}}$ is the superconducting order parameter.

The second term in Eq. (2) represents the long-range Coulomb screening effect. This term vanishes in all symmetries but the completely symmetric one ($A_{1g}$ symmetry). Thus, the Coulomb screening modifies the Raman response only in the totally symmetric case and does not play any role in other symmetries. If we assume a constant $A_{1g}$ vertex function, for example, the second term on the right-hand side of Eq. (2) becomes identical to the first term so that the total response function vanishes (complete screening).

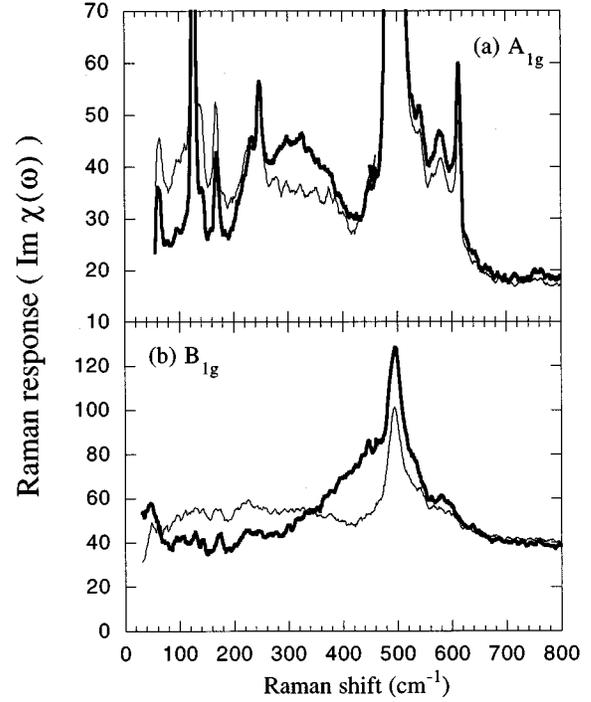

FIG. 1. Raman response function in (a) $A_{1g}$ and (b) $B_{1g}$ symmetries from Tl-2201 (single layer). Thick lines denote spectra taken at 4 K and thin lines at 90 K, which is just above the critical temperature.

Electronic Raman-scattering spectra from Tl-2201 are shown in Fig. 1. Spectra shown here were taken at 4 K (thick lines) and at 90 K (thin lines), which is just above the critical temperature, with a blue excitation (476 nm). Several sharp peaks in the $A_{1g}$ spectra and the peak around 500 cm$^{-1}$ in the $B_{1g}$ spectra are from phononic scattering and have been studied elsewhere.[11,12] Phononic scattering itself is an interesting topic, but strong phonon peaks prevent one from observing the pure electronic scattering. Clear redistribution of the continuum and the appearance of the $2\Delta$ peak are observed in both $A_{1g}$ and $B_{1g}$ spectra. The $2\Delta$ peak positions are measured to be around 320 cm$^{-1}$ in $A_{1g}$ symmetry and 470 cm$^{-1}$ in $B_{1g}$, which gives $2\Delta/k_BT_c$ values of 5.4 for $A_{1g}$ and 8.0 for $B_{1g}$. This is consistent with the results reported from similar samples.[8,11,13,14]

Figure 2 shows $A_{1g}$ and $B_{1g}$ Raman spectra taken from Tl-2212. The spectra shown here are taken with low-energy (red) excitation (647 nm). Low-energy excitation is used primarily to reduce the intensity of phononic scattering which is reported to be very weak with red excitations.[8] Spectra were also taken with the same blue excitation (476 nm), and they show essentially the same characteristics as the spectra taken with the red excitation except for phonon intensities. The $A_{1g}$ and $B_{1g}$ $2\Delta$ peak position in this double CuO$_2$ layer sample are around 430 and 720 cm$^{-1}$, respectively. This gives $2\Delta/k_BT_c$ values of 6.1 ($A_{1g}$) and 10.2 ($B_{1g}$), which are significantly larger than the values from Tl-2201. The large $2\Delta/k_B T_c$ values in Tl-2212 may indicate that this sample is slightly underdoped. The change of $2\Delta/k_B T_c$ ratio by doping has been reported by several authors.[8,13–15] However, we do not yet have a clear understanding of this behavior.

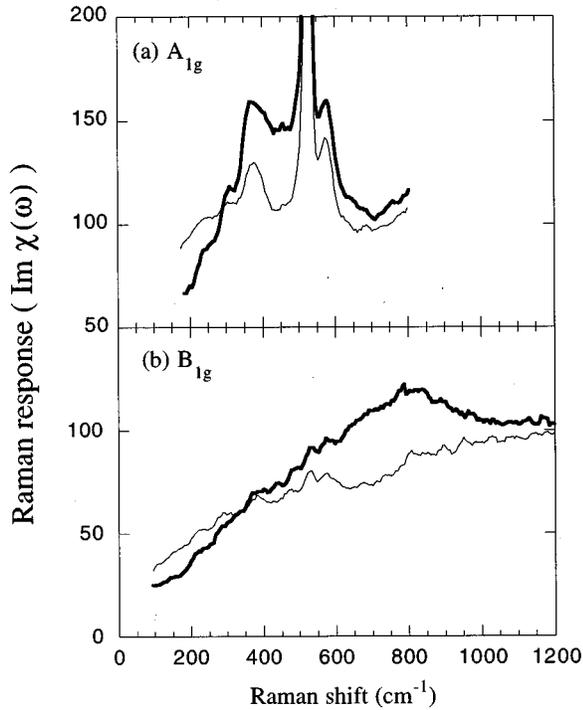

FIG. 2. Raman response function in (a) $A_{1g}$ and (b) $B_{1g}$ symmetries from Tl-2212 (double layer). Thick lines denote spectra taken at 4 K and thin lines at 120 K, which is above the critical temperature.

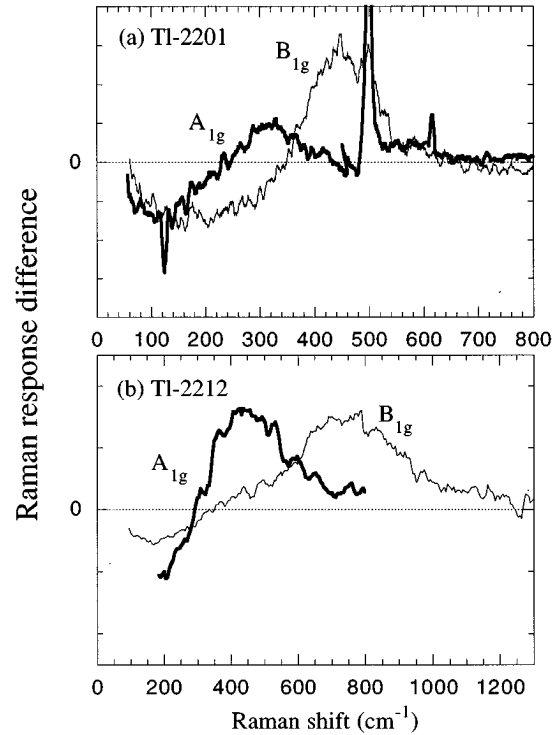

FIG. 3. Differences of Raman response functions in superconducting and normal states in (a) Tl-2201 (single layer) and (b) Tl-2212 (double layer).

To observe suppression of the continuum and appearance of the $2\Delta$ peak better, we subtracted the spectra taken at temperatures above the $T_c$ from the spectra at 4 K. The differences in Raman response (referred to as *difference spectra* hereafter) are presented in Fig. 3. Areas above zero represent the intensity gain in the superconducting state ($2\Delta$ peak), and areas below zero show the suppression of scattering at low frequency due to the opening of the gap. From both samples, the difference becomes zero at sufficiently high frequencies, showing that the superconducting transition does not affect Raman spectra at high frequencies. Sharp features in difference spectra from Tl-2201 are due to temperature dependence of phononic scatterings. From these difference spectra, we measured the $2\Delta$ peak intensity by integrating the area above zero. The continuum intensities are determined by values at sufficiently high frequencies, well above the peak positions where the spectra are flat and are the same above and below the critical temperature. We further normalized the $A_{1g}$ continuum and $2\Delta$ peak intensities with respect to $B_{1g}$ intensities to compare the relative intensities of the $A_{1g}$ $2\Delta$ peak and continuum between Tl-2201 and Tl-2212 as shown in Table I. It is clear that the $A_{1g}$ $2\Delta$ peak is much weaker than the $B_{1g}$ peak in Tl-2201. In the Tl-2212 sample, however, the $A_{1g}$ peak intensity is at least comparable to the $B_{1g}$ intensity. This kind of tendency is also observed in the case of the continuum intensities. Weak $A_{1g}$ $2\Delta$ peak and con-tinuum intensities in Tl-2201 can be explained as a result of the strong Coulomb screening effect in a single $CuO_2$ layer material. The relatively stronger intensities in $A_{1g}$ spectra from Tl-2212 show that the Coulomb screening is not as effective in a double $CuO_2$ layer material. This can be explained in models that extend Eqs. (2)–(4) by adding double bands and Fermi surfaces (even and odd under reflection) in the presence of an interlayer coupling.[16–18] The results of theoretical work based on this idea,[16–18] however, strongly depend on the parameters involved in the models. In reality, the quasiparticle energy has a broad ''tail'' due to strong inelastic scattering among quasiparticles (damping), and the even and odd bands may not be well separated. We believe that a more elaborate description, including the screening effect, interlayer coupling, and the inelastic scattering of quasiparticles, is necessary for a proper explanation of the observed intensity changes.

The ratio of normalized $A_{1g}$ $2\Delta$ peak and continuum intensities from Tl-2201 to those from Tl-2212 are shown in the last row of Table I. These ratios show that the $A_{1g}$ $2\Delta$ peak and continuum intensities in single layer material are only 46–47 % of those in double layer material. These values indicates that the $A_{1g}$ $2\Delta$ peak and the continuum are screened in the same way and affected the same amount by a change in screening. Upon a close examination of the screening term in Eq. (2), it is natural to conclude that electronic Raman scattering for the continuum and the $2\Delta$ peak have a common vertex function or at least vertex functions with very similar **k** dependence. This result is also consistent with a recent resonance study of the $2\Delta$ peak and the continuum in Tl-2201 which shows the vertex functions for the two excitations to have the same excitation energy dependence.[8] The Raman-scattering vertex function describes how photons are coupled to a particular excitation. Thus, similarities in vertex functions for the continuum and the $2\Delta$ peak can be strong evidence that both features have the same origin, and that the $2\Delta$ peak comes from a redistribution of the continuum. This supports the assumption in prior work that the same vertex yields both the $2\Delta$ peak and the continuum.[19,20]

TABLE I. Normalized $A_{1g}$ $2\Delta$ peak and continuum intensities with respect to those in $B_{1g}$. The last row shows the ratio of the normalized $A_{1g}$ $2\Delta$ peak and continuum intensities between Tl-2201 (single layer) and Tl-2212 (double layer).

| | $2\Delta$ peak ($A_{1g}/B_{1g}$) | Continuum ($A_{1g}/B_{1g}$) |
|---|---|---|
| Tl-2201 | 0.33 | 0.44 |
| Tl-2212 | 0.71 | 0.94 |
| Tl-2201/Tl-2212 | 0.46 | 0.47 |

In summary, the superconducting gap measured from the $2\Delta$ peak position in electronic Raman spectra shows the existence of a strongly anisotropic gap in a double $CuO_2$ layer Tl-2212. The $2\Delta/k_B T_c$ values were significantly higher in Tl-2212 than in Tl-2201. The Coulomb screening effect of the $A_{1g}$ excitations is much stronger in single $CuO_2$ layer material than in double $CuO_2$ layer material. Very similar changes in screening effect between single and double layer materials are observed for the $A_{1g}$ $2\Delta$ peak and the $A_{1g}$ continuum, which suggest that the $2\Delta$ peak and the continuum have the same light-scattering mechanism.

This work has been supported by NSF Grant No. DMR 91-20000 through the Science and Technology Center for Superconductivity, and NSF Grant No. DMR 93-20892.